\documentclass[citeautoscript,twocolumn,showpacs,superscriptaddress,prl,aps,floatfix]{revtex4}

\usepackage{graphicx}
\usepackage{rotating}
\usepackage{amsmath}
\usepackage{url}
\usepackage{bm}

\newcommand{\aF}{$\alpha^2 F(\omega)$}

\newcommand{\EF}{$E_F$}
\newcommand{\NEF}{$N(E_F)$}

\newcommand{\mus}{$\mu^*$}
\newcommand{\lam}{$\lambda$}

\newcommand{\tc}{$T_c$}

\begin{document}

\title{Superconductivity in lithium, potassium and aluminium under extreme pressure:\\ A first-principles study}

\author{G.~Profeta}
\affiliation{CASTI - INFM and Dipartimento di Fisica, Universit\`a  dell'Aquila,
I-67010 Coppito (L'Aquila) Italy}

\author{C.~Franchini}
\altaffiliation{Present address: University of Vienna, Institute for Physical Chemistry,
Vienna, Austria.}
\affiliation{SLACS INFM and Dipartimento di Fisica, Universit\`a di Cagliari, I-09042
Monserrato (Ca), Italy}

\author{N.\,N.~Lathiotakis}
\affiliation{Institut f{\"u}r Theoretische Physik, Freie Universit{\"a}t
Berlin, Arnimallee 14, D-14195 Berlin, Germany}

\author{A.~Floris} 
\affiliation{Institut f{\"u}r Theoretische Physik, Freie Universit{\"a}t
Berlin, Arnimallee 14, D-14195 Berlin, Germany}
\affiliation{SLACS INFM and Dipartimento di Fisica, Universit\`a di Cagliari, I-09042 Monserrato (Ca), Italy}

\author{A.~Sanna} 
\affiliation{SLACS INFM and Dipartimento di Fisica, Universit\`a di Cagliari, I-09042
Monserrato (Ca), Italy}

\author{M.\,A.\,L.~Marques}
\affiliation{Institut f{\"u}r Theoretische Physik, Freie Universit{\"a}t
Berlin, Arnimallee 14, D-14195 Berlin, Germany}

\author{M.~L{\"u}ders}
\affiliation{Daresbury Laboratory, Warrington WA4 4AD, United Kingdom }

\author{S.~Massidda} 
\affiliation{SLACS INFM and Dipartimento di Fisica, Universit\`a  di Cagliari, I-09042
Monserrato (Ca), Italy}

\author{E.\,K.\,U.~Gross}
\affiliation{Institut f{\"u}r Theoretische Physik, Freie Universit{\"a}t
Berlin, Arnimallee 14, D-14195 Berlin, Germany}

\author{A.~Continenza}
\affiliation{CASTI - INFM and Dipartimento di Fisica, Universit\`a  dell'Aquila,
I-67010 Coppito (L'Aquila) Italy}

\begin{abstract}
  Extreme pressure strongly affects the superconducting properties of ``simple''
  elemental metals, like Li, K and Al. Pressure induces superconductivity in Li (as high as 17~K), while suppressing it in Al. We report first-principles
  investigations of the superconducting properties of dense Li, K and Al based on a
  recently proposed, parameter-free, method. Our results show an unprecedented
  agreement with experiments, assess the predictive power of the method over
  a wide range of densities and electron-phonon couplings, and provide predictions for K,
  where no experiments exist so far.
   More importantly, our
  results help uncovering the physics of the different behaviors of
  Li and Al in terms of phonon softening and Fermi surface nesting in Li.
\end{abstract}
\pacs{74.25.Jb, 74.25.Kc, 74.20.-z, 74.70.Ad, 71.15.Mb, 74.62.Fj}
\maketitle

The effect of high pressure on phonon mediated superconductors has been the
subject of many investigations. These studies revealed a strong material
dependence: while applied pressure suppresses superconductivity in some
materials, it favors it in others~\cite{ash}.  Even in simple metals, the
physics underlying pressure effects on the superconducting properties can be
very complicated.  For example,
Li~\cite{ash,hanf,lin,shimizu,struz,dee,christ,neaton,liu} and
Al~\cite{gub,sun,daco,bauer}, behave in many circumstances like nearly free-electron
gases, but they exhibit very different behaviors under pressure, still only
partially understood within the Eliashberg theory~\cite{elias}.  At ambient
pressure, Al is a superconductor with $T_c=1.18$~K~\cite{gub}.  
Pressure rapidly reduces \tc\  bringing it down to 0.075~K at
6.2~GPa~\cite{gub}.  Lithium, on the other hand, is a rather complex material:
below 77~K and at zero pressure, it shows a martensitic transition to
energetically competing closed packed structures~\cite{marte}; from 7.5 to
70~GPa it undergoes several structural transitions~\cite{hanf} which suggest the
presence of strong electron-phonon (e--ph) interactions.  No sign of a
transition to a superconducting state above 4~K was found up to $\approx$ 20~GPa
while, at higher pressures, Li becomes a superconductor
~\cite{lin,shimizu,struz,dee}.  In the range 20--38.3~GPa, where Li crystallizes
in an fcc structure, experiments by Shimizu~\cite{shimizu},
Struzhkin~\cite{struz}, and Deemyad~\cite{dee} found that \tc\ increases rapidly
with pressure, reaching values around 12--17~K (the highest \tc\ observed so far
in any elemental superconductor). Also K undergoes several phase transitions, and
is stable in the fcc phase between  11.6  and 23 GPa\cite{EXPK}.

The pressure dependence of \tc\ for Al, Li and K has been calculated by Dacorogna
{\em et al.}~\cite{daco}, by Christensen and Novikov~\cite{christ,christ2}, and by Shi {\em et al.}~\cite{shi} respectively. For Al, Dacorogna {\em et al.}~\cite{daco} 
obtained a nearly satisfactory agreement
with experiments~\cite{gub,sun}. 
In the case of Li, Christensen {\em et al.}~\cite{christ}
used a rigid-muffin-tin approximation for the e--ph coupling constant $\lambda$ and  $\mu^{*}$=0.13.  
Due to the empirical scaling of phonon frequencies \cite{christ2},
they obtained a much too high \tc\ 
(45--75~K) unless the ``non-standard'' value of \mus$\approx 0.22$ or an
additional term modeling spin-fluctuations was used. For K, Shi {\em et al.}~\cite{shi} obtained \tc\ = 9~K at  13.5~ GPa, using $\mu^{*}$=0.13. 

Note that in these reports, as in most other Eliashberg-based
calculations, the electron-electron (e--e) repulsion was treated
semi-empirically through the Morel-Anderson pseudopotential \mus. The validity
of this procedure for the case of low density solids like Li has been questioned
by Richardson and Ashcroft~\cite{ashli}.  Building on the seminal work of
Oliveira, Gross and Kohn~\cite{OGK}, and on further
developments~\cite{kurth}, some of us recently introduced an alternative
approach to Eliashberg theory: an extension of the density functional theory to the
superconducting state (SCDFT)~\cite{noiI}.  This theory is fully {\it ab
  initio}, and is capable of describing correctly the superconducting properties
of several elements~\cite{noiII} and compounds~\cite{noimgb2}. In the present
paper we explore this very promising method to study the superconducting
properties of Li, K and Al under pressure.  Furthermore, we provide a detailed
description of the subtlety of superconductivity in Li, where the incipient
phase transitions produce a phonon softening and a very strong electron-phonon
coupling, thus enhancing \tc\ up to values unusually large for simple elemental
metals.  Our results for Li and K confirm that a full treatment of electronic and phononic
energy scales is required, in agreement with previous arguments~\cite{ashli}.
We predict K to be superconducting, with a \tc\ up to $\approx 2$ K in the experimental
stability range of the fcc structure, and up to  $\approx 11$ K in the range of stable calculated phonon
frequencies.

\begin{figure}[t]
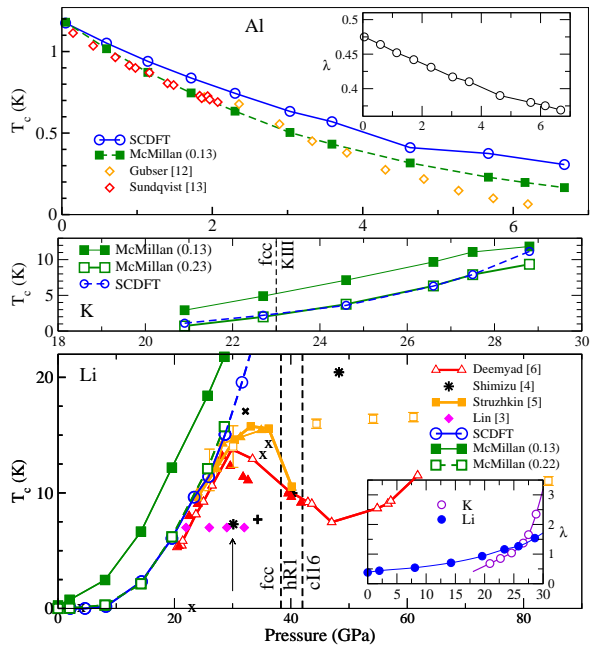

  \begin{center}
    \includegraphics[clip,width=0.43\textwidth]{fig1_a.eps}
    \vspace{0.001cm}
    \includegraphics[clip,width=0.43\textwidth]{fig1_b.eps}
  \end{center}

  \caption{%
    Comparison between calculated and experimental critical temperatures for
    fcc-Al (upper panel), K (middle panel) and fcc-Li (lower panel). Numbers in parenthesis 
    after 'McMillan' indicate the \mus~ value. Different symbols with the same 
    color refer to the same experimental report using different setups. 
    Vertical dashed lines indicate the experimental structural transition 
    pressures for Li and K.  
The insets depict  $\lambda$ {\it vs} pressure in GPa.}
  \label{expTc}
\end{figure}
Ground-state calculations were performed using the pseudopotential based code
PWSCF~\cite{pwscf} within the local density approximation (LDA) to the density
functional theory. The validity of the pseudopotential approach at high
pressures was verified by comparison with all-electron methods. Phonon
frequencies and e--ph couplings were obtained from density functional
perturbation theory. The electron-phonon coupling spectral
function \aF\ and ${\bm q}$-dependent phonon linewidth were evaluated through a
careful integration over the Fermi surface. The implementation of SCDFT
has been reported elsewhere~\cite{noiI,noiII}. All systems were considered in
the fcc structure in their experimental ranges of  stability.

The LDA underestimates the equilibrium volume of
Li and Al, as it can be observed by calculating the equation of state
$P(V)$ using Murnaghan's formula. To compensate for this systematic error, we
apply a positive pressure shift of about $3.5$ and $2$~GPa for Al and Li,
respectively. This is the amount required to match the experimental equation of
state~\cite{hanf,aleos,eqli}. Note that these shifts will always be included in
the values given in the following. No shift is necessary for K.

\begin{figure}[t]
  \begin{center}
    \includegraphics[clip,width=0.43\textwidth]{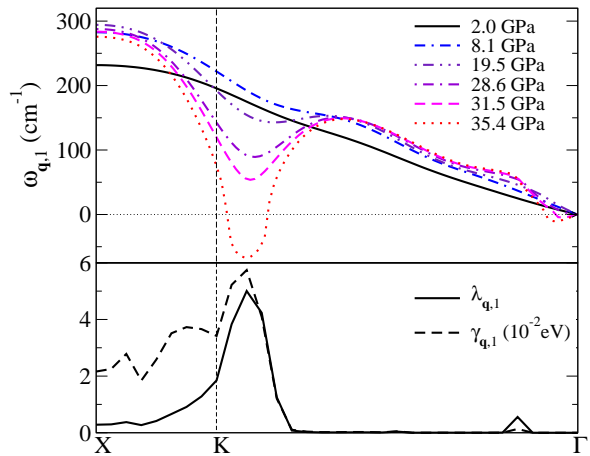}
  \end{center}
  \caption{%
    Upper panel: phonon dispersion of Li along the $X-K-\Gamma$ line, at 
    different pressures, for the lower frequency mode (frequencies below the
    zero axis denote imaginary values).  Lower panel: electron-phonon coupling
    $\lambda_{{\bm q},1}$ and phonon line-width $\gamma_{{\bm q},1}$.}
  \label{soft}
\end{figure}

In Fig.~\ref{expTc}, we compare the calculated pressure dependence of \tc\ for Al, K 
and Li with available experimental results.  For Al, SCDFT calculations match
exactly the experimental zero pressure \tc\ =1.18~K, and reproduce the rapid
decrease of the transition temperature. 
In the same figure we report the
estimation of \tc\ by means of McMillan's formula (using \mus=0.13, in agreement
with previous studies~\cite{daco}). 

\begin{figure}[t]
  \begin{center}
    \includegraphics[clip,width=0.48\textwidth]{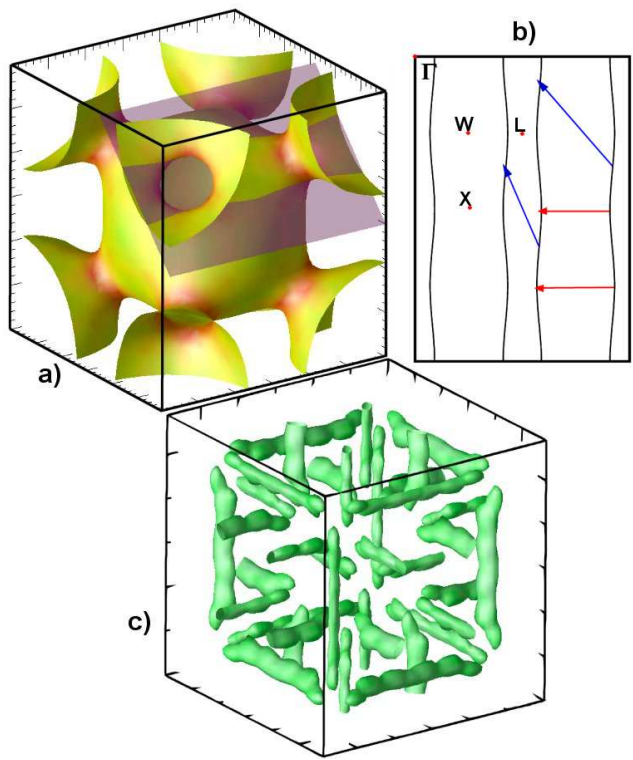}
  \end{center}
  \caption{
    Panel~(a): 3D view of the Fermi surface of Li, at 28.6~GPa, with a color scale
    indicating the value of $\Delta_{\bm k}$.  The red color indicates high
    values of $\Delta$.  Panel~(b): the FS cut on the plane of  Panel~(a), including the 
    {\em L}-point
    and parallel to the (110) plane brings into
    evidence the FS nesting.  The arrows represent nesting vectors.  Panel~(c):
    isosurface of $\lambda_{q,\nu=1}=$~5.  $\Gamma$ is at the vertices and at the center of the cube,
    $X$ at the centers of the faces.}
  \label{Fermi}
\end{figure}

In the case of Li (lower panel of Fig.~\ref{expTc}), despite the
poor agreement among the four sets of experimental data, the most recent
experiments~\cite{dee,struz} agree in: (i)~Li is not superconducting at
ambient pressure; (ii)~$T_c$ is lower than 4~K up to 20~GPa; (iii)~$T_c$ then
increases with pressure, reaching 14~K at about 30~GPa. The only exception to
this behavior is the early report by Lin and Dunn~\cite{lin}. In this pressure
range, the results of Struzhkin~\cite{struz} and of Deemyad~\cite{dee} are quite
similar, while Shimizu {\em et al.}~\cite{shimizu} find lower values of \tc. At
even higher pressures, experiments show a quite complex behavior (see below).
Within this scenario, our calculated SCDFT results are in excellent
quantitative agreement with the most recent experiments~\cite{struz,dee} up to about
30~GPa. We find that Li is not superconducting up to 8~GPa, and that $T_c$ shows
two different trends with pressure, a first region (8--20~GPa) in which $T_c$
increases at a rate of $\sim0.3$~K/GPa, and a second region (20--30~GPa) at
$\sim1.3$~K/GPa.

K shows a behavior quite similar to Li: beyond a pressure threshold (20 GPa) \tc\ rises rapidly. 
In the range where phonons were found to be stable, it reaches $\approx 11$ K at 29~GPa; the
experimentally observed instability of the fcc phase, however, limits this value 
to $\approx 2$ K at 23~GPa. 

The differences between Al and Li can be understood by looking at 
the e--ph coupling as a function of pressure. In Al the phonon 
frequencies increase as the pressure rises (this corresponds to the normal
stiffening of phonons with increased pressure). In addition, the Eliashberg spectral
function  \aF\ shows a high-frequency peak whose  height decreases as a 
function of pressure. These factors
contribute to the decrease of \lam\ (see the inset of Fig~\ref{expTc}) and
consequently of the critical temperature \tc.

In Li, on the
other hand, the phonon frequencies exhibit a quite different behavior. 
In Fig.~\ref{soft} we present
the phonon dispersion of the lowest branch along the $X-K-\Gamma$ line of the
Brillouin zone (BZ). For pressures up to 8~GPa, there is an increase of the
phonon frequencies, as in the case of Al.  However, as the pressure is raised
further, the phonons near the $K$-point start to soften. The softening continues
up to 33~GPa when this frequency becomes imaginary. A closer inspection of
Fig.~\ref{soft} reveals that already at around 30~GPa a phonon mode close to
the $\Gamma$-point develops an imaginary frequency. We believe that this marks
the transition to the hR1 phase, but further analysis would be required to
fully validate this assumption.
Although diffraction experiments performed at 180~K~\cite{hanf} set the
structural phase transition at 39~GPa (up to 42~GPa), it was observed recently that \tc\ 
has a maximum at 30--33~GPa and drops drastically beyond that
pressure~\cite{dee,struz}.  This latter finding, so far unexplained, is
consistent with our theoretical prediction of a complete phonon softening at
around 30~GPa.  

In Fig.~\ref{soft} we plot for the lowest frequency phonon the linewidth
$\gamma_{q,\nu}$ [the Fermi surface (FS) average of the e--ph coupling matrix
elements] and the corresponding ${\bm q}-$dependent electron-phonon coupling
$\lambda_{q,\nu}$, at $P=28.6$~GPa. The quantity $\gamma_{q,\nu}$ shows a peak
close to $K$, and a broad maximum between $K$ and $X$. This peak suggests the
presence of FS nestings. To demonstrate this idea, we plot in Fig.~\ref{Fermi}
a cut of the FS parallel to the (110) plane which includes the $L$ point.  We
clearly recognize nesting vectors (indicated by arrows) connecting fairly flat
and parallel lines. The effects of nesting are remarkably enhanced by the strong e--ph matrix
elements. In  Fig.~\ref{Fermi}(c), we show the isosurface $\lambda_{q,\nu=1}=5$. As it can
be seen, the 
extremely high-coupling regions form tubular structures oriented parallel to the
Cartesian axes, and centered around ${\bm q}$--values along the (110) direction,
matching the FS nesting vectors indicated above. The observed phonon softening
as a function of pressure is a direct consequence of the presence of the FS nesting.  
 In turn,
the progressive FS nesting with the increase of pressure is a consequence of 
the FS topological transition from a spherical free electron-like to a distorted
anisotropic shape (see Fig.~\ref{Fermi}). This topological transition
is a manifestation of an ``$s-p$''
transition of the electronic states near \EF~\cite{hanf,christ, ash2}.  
In particular, 
 we can see in Fig.~\ref{Fermi}(a) that the k-resolved  $\Delta_{\bm k}$ is maximum on
the rings of the FS around the {\em L}, arising from mostly $p-$like and strongly covalently bonded
states. 
An analogous ``$s-d$'' transition occurs in K.
Also the electron localization function (ELF), which
progressively increases with pressure, indicates the departure
from the free-electron picture towards more ``covalent''
phases. 
This produces strong electron-phonon coupling that leads to the
symmetry-breaking phase transitions at high pressure~\cite{hanf} and to the
``paired bonded'' structures close to 100~GPa~\cite{ash2}.

Our calculation of \aF\ for Li leads to a value of $\lambda=0.38$. Using this number and the standard value \mus=0.13 inside McMillan's formula we obtain \tc=0.25~K at
zero pressure, which is lower than the previously reported
one~\cite{christ}, but in agreement with more recent
calculations~\cite{christ2}.  However, it is in complete disagreement with our SCDFT
results and with experiments, both giving \tc=0~K. In order to obtain this
latter result from McMillan's equation we need to use \mus~$\approx 0.22$. This
value also describes quite well the behavior of \tc\ with pressure.
The same physical result was found for K (Fig.~\ref{expTc}), where a very similar value \mus~$\approx 0.23$, 
has to be used to bring McMillan and SCDFT results 
into agreement for all pressures \footnote{At 29~GPa, the value of $\lambda \approx 2$ is out of the range of validity of the McMillan equation. This fact explains the slight difference between the SCDFT and the \mus~$\approx 0.23$ McMillan results at that pressure.}.
 According to Richardson and Ashcroft~\cite{ashli}, a large value of
\mus\ is justified by the fact that the e--e interaction becomes unusually large
at such densities. These authors pointed out that, in this density range, a full
treatment of electrons and ions on the same footing is required. This is exactly what our method ~\cite{noiI, noiII} achieves: The different energy scales of
Coulomb repulsion and phonon-mediated attraction are fully included without any {\it ad-hoc} modeling.

In order to further investigate the effect of the Coulomb repulsion, we solved
the SCDFT gap equation in Li using for the e--e repulsion both the actual matrix
elements of the Thomas-Fermi screened potential and the approximate functional
$\cal K^\text{TF-SK}$ introduced in Ref.~\cite{noiII}.  While this latter
approximation failed for the localized $\sigma$ orbitals in MgB$_2$, it works
very well for Li over the pressure range considered. Thus, while the strong
e--ph coupling and the ELF indicate that covalency is becoming progressively
more important, the Coulomb repulsion can still be modeled as if electrons were
delocalized. Then, why \mus\ has to be so large? We can try to answer this question using
SCDFT. By making a comparative analysis between Al and Li we found that the FS
average of the Coulomb potential multiplied by \NEF\ (which can be directly
related to $\mu$) is $\approx 40$\% larger in Li than in Al, mostly because of
the Van Hove peak in the density of states of Li. This effect is present even
after taking into account the renormalization of the Coulomb repulsion due to
retardation effects. Thus, our results clearly demonstrate a larger value of the
Coulomb repulsion in Li than in Al.  Since our functional, based on the
Thomas-Fermi screened Coulomb repulsion, does not include spin-fluctuations, we
can clearly rule out that spin-fluctuations are required to explain the large Coulomb
contributions and the anomalously high value of \mus.  Further
investigations at $T=0$ indicate that the gap function is very sensitive to
changes of the density of states at all energies, once again showing the
difficulty to reduce the complexity of the Coulomb interaction into a single
numerical parameter.
 
In summary, the recently introduced SCDFT method allowed us to calculate the
superconducting transition temperature of Al, K and Li under high pressure from
first principles. The results obtained for Al and Li are in very good agreement with
experiment, and account for the opposite behavior of these two metals under
pressure. Furthermore, the increase of \tc\ with pressure in Li is explained in
terms of the strong e--ph coupling, which is due to changes in the topology of
the Fermi surface, and is responsible for the observed structural instability.
Finally, our results for K provide predictions intriguing enough 
to suggest experimental work on this system.

Note: during the reviewing process of this paper two reports appeared confirming
phonon softening~\cite{JPC} and Fermi surface deformation~\cite{prieto} in Li
under pressure.

\end{document}